\documentclass[useAMS,usenatbib,epsfig]{mn2e}
\usepackage{natbib}
\usepackage{amsmath,amsfonts}
\usepackage{epsfig}
\def\reff@jnl#1{{\rm#1\/}}

\def\aj{\reff@jnl{AJ}}                  % Astronomical Journal
\def\araa{\reff@jnl{ARA\&A}}            % Annual Review of Astron and Astrophys
\def\apj{\reff@jnl{ApJ}}                        % Astrophysical Journal
\def\apjl{\reff@jnl{ApJ}}               % Astrophysical Journal, Letters
\def\apjs{\reff@jnl{ApJS}}              % Astrophysical Journal, Supplement
\def\ao{\reff@jnl{Appl.Optics}}         % Applied Optics
\def\apss{\reff@jnl{Ap\&SS}}            % Astrophysics and Space Science
\def\aap{\reff@jnl{A\&A}}               % Astronomy and Astrophysics
\def\aapr{\reff@jnl{A\&A~Rev.}}         % Astronomy and Astrophysics Reviews
\def\aaps{\reff@jnl{A\&AS}}             % Astronomy and Astrophysics, Supplement
\def\azh{\reff@jnl{AZh}}                        % Astronomicheskii Zhurnal
\def\baas{\reff@jnl{BAAS}}              % Bulletin of the AAS
\def\jrasc{\reff@jnl{JRASC}}            % Journal of the RAS of Canada
\def\memras{\reff@jnl{MmRAS}}           % Memoirs of the RAS
\def\mnras{\reff@jnl{MNRAS}}            % Monthly Notices of the RAS
\def\pra{\reff@jnl{Phys.Rev.A}}         % Physical Review A: General Physics
\def\prb{\reff@jnl{Phys.Rev.B}}         % Physical Review B: Solid State
\def\prc{\reff@jnl{Phys.Rev.C}}         % Physical Review C
\def\prd{\reff@jnl{Phys.Rev.D}}         % Physical Review D
\def\prl{\reff@jnl{Phys.Rev.Lett}}      % Physical Review Letters
\def\pasp{\reff@jnl{PASP}}              % Publications of the ASP
\def\pasj{\reff@jnl{PASJ}}              % Publications of the ASJ
\def\qjras{\reff@jnl{QJRAS}}            % Quarterly Journal of the RAS
\def\skytel{\reff@jnl{S\&T}}            % Sky and Telescope
\def\solphys{\reff@jnl{Solar~Phys.}}    % Solar Physics
\def\sovast{\reff@jnl{Soviet~Ast.}}     % Soviet Astronomy
\def\ssr{\reff@jnl{Space~Sci.Rev.}}     % Space Science Reviews
\def\zap{\reff@jnl{ZAp}}                        % Zeitschrift fuer Astrophysik
\def\nat{\reff@jnl{Nature}}             % Nature 

\title[Cosmological parameter estimation using VSA data]{Cosmological
parameter estimation using Very Small Array data out to $\ell=1500$}
\label{firstpage}
\author[R. Rebolo et al.] 
{Rafael Rebolo$^{1*}$, 
 Richard A. Battye$^{2\dag}$,
 Pedro Carreira$^2$, 
 Kieran Cleary$^2$, 
 Rod D. Davies$^2$, 
\newauthor 
 Richard J. Davis$^2$,  
 Clive Dickinson$^2$, 
 Ricardo Genova-Santos$^1$,
 Keith Grainge$^3$, 
\newauthor
 Carlos M. Guti{\'e}rrez$^1$,  
 Yaser A. Hafez$^2$,
 Michael P. Hobson$^3$,  
 Michael E. Jones$^3$, 
\newauthor
 R\"udiger Kneissl$^3$, 
 Katy Lancaster$^3$, 
 Anthony Lasenby$^3$,  
 J. P. Leahy$^2$,
\newauthor 
 Klaus Maisinger$^3$,  
 Guy G. Pooley$^3$, 
 Nutan Rajguru$^3$, 
Jos\'e Alberto Rubi\~no-Martin$^{1,4}$,
\newauthor
Richard D.E. Saunders$^3$, 
Richard S. Savage$^{3\ddag}$,  
Anna Scaife$^3$,
Paul F. Scott$^3$,  
\newauthor
An\v ze Slosar$^{3\S}$,
Pedro Sosa Molina$^1$,
 Angela C. Taylor$^3$,  
 David Titterington$^3$,
\newauthor  
 Elizabeth Waldram$^3$,  
 Robert A. Watson$^{2\P}$,
Althea Wilkinson$^2$
\\
  $^1$Instituto de Astrof{\'{\i}}sica de Canarias, 38200 La Laguna,
  Tenerife, Spain.\\ 
  $^2$University of Manchester, Jodrell Bank Observatory, UK.\\
  $^3$Astrophysics Group, Cavendish Laboratory, University of Cambridge, UK.\\
  $^{4}$Present address: Max-Planck Institut f\"ur Asrtrophysik, Karl-Schwarzchild-Str. 1, Postfach 1317, 85741, Garching, Germany.\\
  $^{\ddag}$Present address: Astronomy Centre, University of Sussex, UK.\\
  $^{\S}$Present address: Faculty of Mathematics \& Physics, University of Ljubljana, 1000 Ljubljana, Slovenia.\\
  $^{\P}$Present address: Instituto de Astrof{\'{\i}}sica de
Canarias, 38200 La Laguna, Tenerife, Spain.\\
  $^{*}$Email: rll@ll.iac.es\\
  $^{\dag}$Email: rbattye@jb.man.ac.uk\\}
 
\date{Accepted ---; received ---; in original form \today}
\pagerange{\pageref{firstpage}--\pageref{lastpage}}
\pubyear{2003}

\begin{document}
%\label{firstpage}
\maketitle

\begin{abstract}
We estimate cosmological parameters using data obtained by the Very
Small Array (VSA) in its extended configuration, in conjunction with a
variety of other CMB data and external priors. Within the flat
$\Lambda$CDM model, we find that the inclusion of high resolution data
from the VSA modifies the limits on the cosmological parameters as
compared to those suggested by WMAP alone, while still remaining
compatible with their estimates. We find that $\Omega_{\rm b}h^2=0.0234^{+0.0012}_{-0.0014}$, $\Omega_{\rm dm}h^2=0.111^{+0.014}_{-0.016}$, $h=0.73^{+0.09}_{-0.05}$, $n_{\rm S}=0.97^{+0.06}_{-0.03}$, $10^{10}A_{\rm S}=23^{+7}_{-3}$ and $\tau=0.14^{+0.14}_{-0.07}$ for WMAP and VSA when no external prior is included.On extending the model to include
a running spectral index of density fluctuations, we find that the
inclusion of VSA data leads to a negative running at a level of more 
than $95\%$ confidence ($n_{\rm run}=-0.069\pm 0.032$), something which is
not significantly changed by the inclusion of a stringent prior on the Hubble
constant. Inclusion of prior information from the 2dF galaxy redshift
survey reduces the significance of the result by constraining the
value of $\Omega_{\rm m}$. We discuss the veracity of this result in
the context of various systematic effects and also a broken spectral
index model. We also constrain the fraction of neutrinos and find that
$f_{\nu}< 0.087$ at $95\%$ confidence which corresponds to 
$m_\nu<0.32{\rm eV}$ when all neutrino masses are the equal.
Finally, we consider the global
best fit within a general cosmological model with 12 parameters and
find consistency with other analyses available in the literature. The evidence for $n_{\rm run}<0$ is only marginal within this model.
  
\end{abstract}


\section{Introduction}

Recent measurements of the cosmic microwave background (CMB) anisotropies
have allowed the determination of a large number of cosmological
parameters with unprecedented accuracy. The Wilkinson Microwave
Anisotropy Probe (WMAP) and pre-WMAP data sets can be fitted by a
six-parameter $\Lambda$CDM model (see, for example, Bennett et
al. 2003, Slosar et al. 2003).
In order to break the degeneracies inherent in the CMB power
spectrum (Efstathiou \& Bond 1999), various authors have augmented measurements of the CMB 
with observations of large-scale structure (LSS), for example, the 2dF
galaxy power spectrum (Percival et al. 2001, Percival et al. 2002),
 Lyman-$\alpha$ forest (Mandelbaum et al. 2003; Seljak et al. 2003), SDSS
three-dimensional power spectrum (Tegmark et al. 2003), measurements
of cosmic shear (Hoekstra et al. 2002) and the galaxy cluster
luminosity function (Allen et al. 2003a), and/or information on the
expansion rate of the Universe from measurements of the Hubble
constant (Freedman et al. 2001) and high redshift supernovae (Perlmutter et
al. 1999; Reiss et al. 1998).

High-resolution ($\ell\ge 700$) observations of CMB anisotropies provided
by previously released data obtained by the Very Small Array (VSA,
Grainge et al. 2003), the Arcminute Bolometer Array (ACBAR, Kuo et
al. 2004) and the Cosmic Background Imager (CBI, Pearson et al. 2003),
can also be important in reducing the impact of degeneracies and
provide information on the parameters relating to the power spectrum
of initial density fluctuations over a much wider range of scales. In
particular, the WMAP team made use of these data in their
analyses in order to improve the significance of their results (Spergel et al. 2003).

In this paper, we study the cosmological implications of the new CMB
power spectrum measured by the VSA which has a good signal-to-noise
ratio out to a multipole of $\ell=1500$ (Dickinson et al. 2004).
These observations cover 33 fields, as opposed
to 9 in Grainge et al. (2003), representing an improvement of $\sim 2$ in
signal-to-noise over the previous data. By virtue of the accurately
measured temperature of Jupiter by WMAP, the absolute calibration uncertainty
for these data is reduced to $3\%$ on the power spectrum; something
which will be significant in our subsequent discussion. The power
spectrum is measured between $\ell=300$ and $\ell=1500$ with a
resolution in $\ell$-space of $\Delta\ell\approx60$. Previous measurements
of the power spectrum between $\ell=130$ and $\ell=900$
using the VSA compact configuration can be found in Scott et al. (2003).

We will first consider the standard six-parameter flat $\Lambda$CDM
model, and then include extra parameters broadly in keeping with the
approach taken in the papers published by the WMAP team (Spergel et
al. 2003, Verde et al. 2003, Peiris et al. 2003). Our main focus will
be on the initial spectrum of fluctuations, quantified by the running
of the spectral index, which appears to be particularly sensitive to
high-resolution data such as ours. In the case where we do not impose
external priors on the CMB data (WMAP+VSA), we find that there is
significant evidence ($>2\sigma$) for negative running; something which is not
implied by the WMAP data alone. The significance of this result is
sensitive to the inclusion of external priors, the relative
calibration of WMAP and VSA, and possible source/cluster contamination
of the measured power spectrum, illustrating issues which are of great
relevance in the era of precision cosmology. The result, if true,
would be a significant challenge to models of slow-roll inflation, and
so we also consider a broken spectral index model. As a final point,
we consider a 12-parameter model fit to WMAP, WMAP+VSA and all
available CMB data beyond $\ell>1000$, illustrating the effects of
external priors on the estimated parameters. Our results within this
model are compatible with previous determinations, both by the WMAP
team and others.

\section{Methodology}

\subsection{Cosmological model}
\label{sec:cosmomodel}

We will define the $\Lambda$CDM model as follows. First, we
will assume that the Universe is flat and dominated by cold dark
matter (CDM), baryons and a cosmological constant, $\Lambda$. The
densities of these components relative to critical are denoted
$\Omega_{\rm dm}$, $\Omega_{\rm b}$ and $\Omega_\Lambda$ respectively
and we define $\Omega_{\rm m}=\Omega_{\rm dm}+\Omega_{\rm b}$ to be
the overall matter density (CDM and baryons) in the same units. The
expansion rate is quantified in terms of the Hubble constant
$H_{0}=100h\,{\rm km}\,{\rm sec}^{-1}\,{\rm Mpc}^{-1}$ and we allow
for instantaneous reionization at some epoch $z_{\rm re}(<30)$ which
can also be quantified in terms of an optical depth $\tau$. The
so-called physical densities of the CDM and baryons are defined as
$\omega_{\rm dm}=\Omega_{\rm dm}h^2$ and $\omega_{\rm b}=\Omega_{\rm
b}h^2$. We will consider only adiabatic models and, guided by the
predictions of slow-roll inflation, we parameterize the initial
fluctuation spectrum of this model by
\begin{equation} 
P(k)=A_{\rm S}\left({k\over k_{\rm c}}\right)^{n_{\rm S}}\,,
\end{equation}
where $k_{\rm c}=0.05{\rm Mpc}^{-1}$ is the arbitrarily chosen pivot
point of the spectrum, $n_{\rm S}$ is the spectral index and $A_{\rm
S}$ is the scalar power spectrum normalization.

We will modify this model by the inclusion of two other
parameterizations of the power spectrum. We will, for the most part,
consider a model with a running spectral index, 
\begin{equation}
P(k)=A_{\rm S}\left({k\over k_{\rm c}}\right)^{n_{\rm S}+{1\over 2}n_{\rm run}\log(k/k_{\rm c})}\,,
\end{equation}
so that the overall
spectral index of fluctuations is a function of scale, $n_{\rm S}(k)$, 
given by
\begin{equation}
n_{\rm S}(k)={d(\log\,P)\over d(\log\,k)}=
=n_{\rm S}+n_{\rm run}\log\left({k\over k_{\rm c}}\right)\,,
\end{equation}
where $n_{\rm run}$ is  known as the running of the spectral
index. For slow roll inflation to be well defined, one requires that
$|n_{\rm run}|\ll |1-n_{\rm S}|/2$ (Leach and Liddle, 2003).
Under certain choices of priors we
find that there is some evidence that this inequality is violated by
the preferred fits to the data. We therefore consider an alternative
model, which could be motivated by broken-scale invariance models of
inflation (see, for example, Barriga et al. 2001),
but is probably best thought of as a test of
whether or not the data prefer a single power-law. The specific choice
we will make is to consider
\begin{equation}
n(k)=n_1\hbox{ for }k<k_{\rm c}\hbox{ and }n(k)=n_2\hbox{ for }k>k_{\rm c}\,,
\end{equation}
with an appropriate normalization for $k>k_{\rm c}$ so as to make the
power spectrum continuous and the same value of $k_{\rm c}$ as used in
the standard $\Lambda$CDM model. 

In our discussion of systematic effects in section~\ref{sec:disc}, we
will consider the possibility of an extra component to the
anisotropies with $C_{\ell}=2\pi A_{\rm X}\times 10^{-6}$.  Such a
component is motivated by foreground effects due to point sources and
the Sunyaev-Zel'dovich (SZ) effect from galaxy clusters along the line
of sight. The temperature anisotropies due to such a component will be
$(\Delta T_\ell)^2=(\ell(\ell+1)C_{\ell}/(2\pi))=A_{\rm X}\ell^2$ and could be significant for
$\ell>1000$. The VSA has a sophisticated procedure to extract the
effects of point sources using a dedicated, co-located, single-baseline
interferometer (see Dickinson et al. (2004)
for details), and the VSA
fields have been chosen to avoid very luminous X-ray clusters. There
could still, however, be some residual contamination. Moreover, claims
have been made of an excess signal between $\ell=2000$ and $\ell=4000$
by the CBI team (Mason et al. 2003), who attribute this to the SZ
effect. If the signal is a large as is claimed, then it could be a
contaminant even at lower $\ell$. By including such a component in
the parameter fitting, it should be possible to constrain the
contribution at $\ell>2000$ as well as gaining some insights into the
possible systematic effects of making such an error.

The other parameters which we will consider in our analyses are:
$f_{\nu}=\Omega_{\nu}/\Omega_{\rm dm}$, the fraction of the dark matter 
which is massive
neutrinos; $\Omega_{\rm k}=1-\Omega_{\rm tot}$ ($\Omega_{\rm
tot}=\Omega_{\rm dm}+\Omega_{\rm b}+\Omega_{\nu}+\Omega_{\Lambda}$),
the curvature in units of the critical density; $w=P_{\rm Q}/\rho_{\rm
Q}$, the equation-of-state parameter for a dark energy component
modelled as a slowly rolling scalar field; $n_{\rm T}$ the spectral
index of tensor fluctuations specified at the pivot point $k_{\rm
c}=0.002\,{\rm Mpc}^{-1}$; $R=A_{\rm T}/A_{\rm S}$, the ratio of the
amplitude of the scalar fluctuations, $A_{\rm S}$, evaluated at
$k_{\rm c}=0.05\,{\rm Mpc}^{-1}$, and that of the tensor fluctuations
evaluated at $k_{\rm c}=0.002\,{\rm Mpc}^{-1}$.  In addition to these
parameters, for which we fit, we will also comment on various derived
quantities: $t_0$, the age of the universe; $\sigma_{8}$, the
amplitude of density fluctuations in the spheres of $8h^{-1}\,{\rm
Mpc}$.

\subsection{CMB data}

In this paper we will consider the cosmological implications of four
different combinations of CMB data. 
\begin{itemize}
\item The first data set, denoted {\sc
COBE+VSA} contains the VSA data as described in Dickinson et al. (2004)
combined with the COBE data (Smoot et al. 1992, Bennett et al. 1996).
The purpose of this particular
data set is to check the consistency of the VSA data with the
concordant model, without imposing the strong constraining power of
the WMAP data set (Bennett et al. 2003) . 

\item The second data set, denoted
{\sc WMAP} contains only the WMAP temperature (TT) data (Hinshaw et
al. 2003) and temperature-polarization cross-correlation
(TE) data (Kogut et al. 2003). We use these
data sets to provide a meaningful comparison with cosmological results
obtained from other data sets, avoiding differences that might
arise due to the priors and other methodological issues.

\item The third data set contains WMAP data and the new
VSA data and is referred to
as {\sc WMAP+VSA}. In this data-set we supplement the accurate
measurement of the first two peaks by the WMAP satellite with the VSA
measurements of the power spectrum in the region between the third and
fifth peaks. The importance of these data set is to illustrate the
extra information that is available from the measurements of the power
spectrum on small angular scales.

\item The last data set combines the previous two with all important
CMB experiments providing measurements in the region of the second
peak of the spectrum and beyond, namely CBI, ACBAR, Boomerang, Maxima,
DASI (Pearson et al. 2003; Kuo et al. 2004; Netterfield et al. 2002;
Hanany et al. 2002 and Halverson et al. 2002, respectively).
This last data set is hereafter referred to as {\sc AllCMB}.
\end{itemize}

Throughout our analysis we ignore small correlations between data sets
that arise due to the fact that they have observed the same portions of
the sky. This applies only to correlations between WMAP, that has used
nearly all the sky, and terrestrial experiments, that have observed only
small patches. In all cases, the decoupling of observed
angular scales and the fact that any given patch of sky observed by a
terrestrial experiment makes up less than 1\% of the WMAP sky coverage
makes this approximation truly valid and far below systematic
uncertainties.

\subsection{External priors}
\label{extpriors}

In addition to the CMB data sets described above, we consider the
effects of other cosmological data, not only to break the
degeneracies, but also to see how the measured CMB power spectrum fits
in the wider cosmological context. Each of these 
`external priors' is discussed below.

\begin{itemize}

\item The constraint on Hubble's constant obtained by imposing the Hubble
Space Telescope (HST) Key project value of $H_0=72\pm8\,{\rm km}
\,{\rm sec}^{-1}\,{\rm Mpc}^{-1}$ (Freedman et al. 2001) as a Gaussian
distribution. The error-bar includes both statistical and systematic
uncertainty and prohibits the low density, low $h$ universes allowed by
the CMB data alone.

\item Constraints on large scale structure
from the 2dF Galaxy Redshift Survey (Colless et al. 2001; Percival et
al 2001; Percival et al 2002), which provides measurements on scales $0.02 <
k/(h\,{\rm Mpc}^{-1})<0.15$. The 2dF data measure the power spectrum of the
matter fluctuations in the linear regime, which is linked to the spectrum of
primordial fluctuations and the parameters of the standard model in a
different manner to the CMB data and, thus, provide an important
consistency check.

\item Constraints from Type Ia Supernovae (SNeIa) (Perlmutter et al. 1999, Reiss et al. 1998), which
help to break the CMB geometrical degeneracy and thus accurately
determine the ratio of matter to dark-energy components in our
Universe.

\item Constraints from the gas fraction ($f_{gas}$) in dynamically relaxed
clusters of galaxies (Allen et al. 2002) and from the observed local
X-ray luminosity function (XLF) of galaxy clusters (Allen et
al. 2003a). These data provide very accurate measurements of matter
content of our universe, albeit with large systematic uncertainties. 

\item Constraints from cosmic shear (CS) measurements (Hoekstra et
al. 2002), which provide an independent restriction in 
the $\Omega_m$-$\sigma_8$ plane from that implied by X-ray
observations of clusters. 

\end{itemize}

\subsection{Parameter estimation}

The parameter estimation has been performed using the {\sc cosmomc}
computer package (Lewis \& Bridle 2002) using the April 2003 version
of the software (note that the default parametrisation is different in
the more recent versions of the {\sc cosmomc} package). The
calculations were performed on LAM clusters with a total of 42 CPUs
at the IAC in La
Laguna, Tenerife and the {\sc COSMOS} supercomputer facility at the
University of Cambridge.  The {\sc cosmomc} software uses the Markov
Chain Monte Carlo (MCMC) algorithm to explore the hypercube of
parameters on which we impose flat priors. These priors are listed in
Table.~\ref{tab:pri}. Additionally, the software automatically
imposes the physical prior $\Omega_\Lambda>0$, which can significantly
affect the marginalized probability distributions (see Slosar et al. 2003
for further discussion).  For each considered model, we have run the
software until 1 in 25 of samples are accepted. Once this is achieved we ignore the first 200 accepted samples as a burn-in phase. In the flat models this leads to 65000 independent samples, and 200000 in the non-flat case. These samples were then thinned by a factor 25 and 
used to plot marginalized probability distributions with the {\sc
getdist} facility, which is part of the standard {\sc cosmomc}
package. This program uses a smoothing kernel to infer a sufficiently
smooth posterior probability curve from discrete MCMC samples.

\begin{table}
\caption{Priors used on each cosmological parameter when it is allowed
to vary. The notation $(a,b)$ for parameter $x$ denotes a top-hat
prior in the range $a \le x \le b$.}
\begin{center}
\begin{tabular}{lc}
\hline
Basic Parameter & Prior \\
\hline
$\omega_{\rm b}$    & (0.005,0.10) \\
$\omega_{\rm dm}$   & (0.01, 0.99) \\
$h$                 & (0.4,1.0) \\
$n_{\rm S}, n_1, n_2$     & (0.5,1.5) \\
$z_{\rm re}$            & (4,30) \\
$10^{10}A_{\rm S}$        & (10,100)  \\ 
$n_{\rm run}$ & ($-0.15$,0.15)  \\
$A_{\rm X}/(\mu{\rm K})^2$ & ($-500$,500) \\
$f_{\nu}$ & (0,0.2) \\
$\Omega_{\rm k}$  & ($-0.25$,0.25) \\ 
$w$ & ($-1.5$,0) \\ 
$R$ & (0,2) \\ 
$n_{\rm T}$& ($-1.5$,3) \\
\hline
\label{tab:pri}
\end{tabular}
\end{center}
\end{table}

\section{Results}

\subsection{Flat $\Lambda$CDM models}

\subsubsection{Standard six-parameter model}

\begin{figure}
\includegraphics[width=8cm,height=12cm,angle=0]{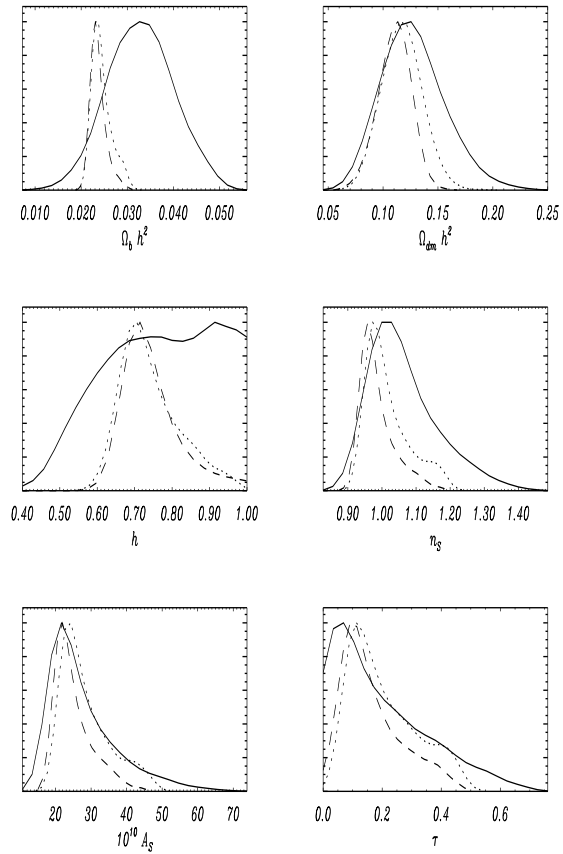}
\caption{Marginalized distributions for the standard 6-parameter flat
$\Lambda$CDM model with no external priors (that is, CMB alone) using
COBE+VSA (solid-line), WMAP alone (dotted line) and WMAP+VSA (dashed
line).}
\label{fig:lcdm}
\end{figure}

\begin{table}
\caption{Parameter estimates and 68\% confidence limits for the
standard six-parameter flat $\Lambda$CDM model.}
\begin{tabular}{lcccc}
\hline
Parameter & COBE+VSA & WMAP & WMAP+VSA\\
\hline \\
$\omega_{\rm b}$    & $0.0328^{+0.0073}_{-0.0071}$& $0.0240^{+0.0027}_{-0.0016}$ &  $0.0234^{+0.0019}_{-0.0014} $\\ \\
$\omega_{\rm dm}$   & $0.125^{+0.031}_{-0.027}$& $0.117^{+0.018}_{-0.018}$   &  $0.111^{+0.014}_{-0.016}$ \\ \\
$h$                 & $0.77^{+0.15}_{-0.17}$& $0.73^{+0.10}_{-0.06}$ &  $0.73^{+0.09}_{-0.05}$ \\ \\
$n_{\rm S}$         & $1.05^{+0.12}_{-0.08}$& $1.00^{+0.09}_{-0.04}$  & $0.97^{+0.06}_{-0.03}$\\ \\
$10^{10}A_{\rm S}$  &  $25^{+11}_{-6}$& $27^{+9}_{-5}$ &   $23^{+7}_{-3}$\\ \\
$\tau$              & Unconstrained& $0.18^{+0.16}_{-0.08}$  &  $0.14^{+0.14}_{-0.07}$\\ \\
\hline
\label{tab:lcdm}
\end{tabular}
\end{table}

We begin our discussion in the context of the standard
flat $\Lambda$CDM model with six free parameters ($\omega_{\rm
b}$, $\omega_{\rm dm}$, $h$, $n_{\rm S}$, $A_{\rm S}$, $\tau$), which was
discussed in Spergel et al. (2003) for WMAP, with no external priors.
We should note that it is, in fact, $z_{\rm re}$ which we allow to
vary in our analysis, but we present $\tau$ to be consistent with 
previous work.
 
The marginalized distributions for the parameters are presented in
Fig.~\ref{fig:lcdm} and the derived parameter estimates are tabulated
in Table~\ref{tab:lcdm}. The values for WMAP alone can be compared
with those in Spergel et al. (2003). Noting that they present
$\omega_{\rm m}=\Omega_{\rm m}h^2$, instead of $\omega_{\rm dm}$,
there are only minor discrepancies in the central values, although some 
of the limits appear to be somewhat larger. The
preferred value of the redshift of reionization is $z_{\rm
re}=17^{+8}_{-6}$. The inclusion of the high-resolution data from the
VSA modifies the limits on each of the parameters as one can see from
Fig.~\ref{fig:lcdm} and these are most significant for $n_{\rm S}$,
whose best fitting value reduces from 1.00 to 0.97.
The result for $n_{\rm S}$ will be
central to our subsequent discussion of the primordial power spectrum. The
results from WMAP+VSA are very similar to those presented in Spergel
et al. (2003) for WMAP+ACBAR+CBI.

We have also included in Fig.~\ref{fig:lcdm} the marginalized
distributions and derived limits obtained from the COBE+VSA data set,
all of which show compatibility with the results of WMAP.  One
slightly unusual result is that for $\omega_{\rm b}$ which is much
larger than the value suggested by WMAP, WMAP+VSA and standard Big
Bang Nucleosynthesis, $\omega_{\rm b}=0.020\pm 0.002$, (Burles,
Nollett and Turner 2001) and is a result of somewhat larger amplitude
of the 3rd peak and the shifted first peak preferred by the VSA data (Rubi{\~ n}o-Martin et al. 2003)  in isolation (see Dickinson et al. 
2004 for a detailed discussion of the preferred peak structure of the current data). Comparing the derived distributions with
those obtained by Slosar et al. 2003 (using the earlier VSA data presented in 
Grainge et al. 2003),
we find that the results are fully consistent, but the additional
VSA data has led to tighter parameter constraints. In particular, the
upper limit on $\omega_{\rm dm}$ has been significantly reduced.

\subsubsection{Running spectral index models}

\begin{figure}
\includegraphics[width=8cm,height=12cm,angle=0]{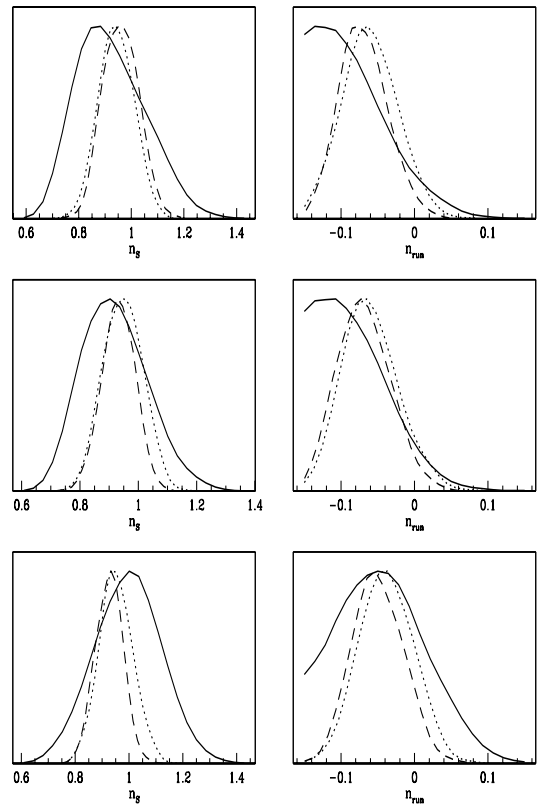}
\caption{Marginalized distributions for $n_{\rm S}$ and $n_{\rm run}$
in the flat $\Lambda$CDM model with a running spectral index. 
Line-styles are as in Fig.~\ref{fig:lcdm}. The external priors
adopted are: none (top row), HST (middle row), 2dF (bottom row).}
\label{fig:nsnrun}
\end{figure}

\begin{figure}
\includegraphics[width=8cm,height=12cm,angle=0]{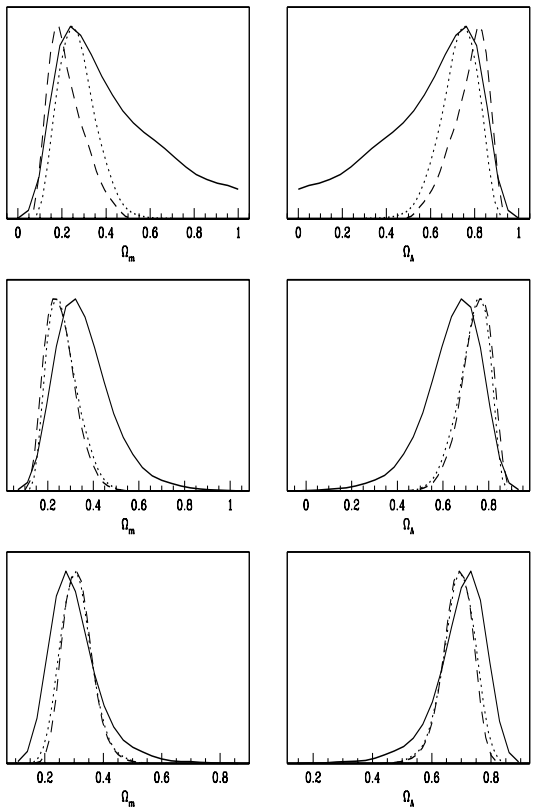}
\caption{As for Fig.~\ref{fig:nsnrun}, but for the parameters
$\Omega_{\rm m}$ and $\Omega_\Lambda$.}
\label{fig:omolam}
\end{figure}

\begin{table}
\caption{Limits on $n_{\rm S}$ and $n_{\rm run}$ in the flat
$\Lambda$CDM model with a running spectral index for different 
CMB data sets and external priors.}
\begin{tabular}{lccc}
\hline
CMB  & External & $n_{\rm S}$ & $n_{\rm run}$ \\
\hline \\
COBE+VSA & None & $0.93^{+0.13}_{-0.12}$ & $-0.081^{+0.049}_{-0.049}$ \\ \\
WMAP     & None & $0.94^{+0.07}_{-0.06}$ & $-0.060^{+0.037}_{-0.036}$ \\ \\
WMAP+VSA & None & $0.96^{+0.07}_{-0.07}$ & $-0.069^{+0.032}_{-0.032}$
\\ \\[3mm]
%\hline \\
COBE+VSA & HST  & $0.92^{+0.11}_{-0.12}$ & $-0.081^{+0.048}_{-0.048}$ \\ \\
WMAP     & HST  & $0.95^{+0.06}_{-0.07}$ & $-0.060^{+0.037}_{-0.037}$ \\ \\
WMAP+VSA & HST  & $0.93^{+0.06}_{-0.05}$ & $-0.069^{+0.036}_{-0.036}$ 
\\ \\[3mm]
%\hline\\
COBE+VSA & 2dF  & $1.00^{+0.12}_{-0.13}$ & $-0.044^{+0.058}_{-0.061}$ \\ \\
WMAP     & 2dF  & $0.95^{+0.05}_{-0.06}$ & $-0.038^{+0.025}_{-0.037}$ \\ \\
WMAP+VSA & 2dF  & $0.93^{+0.05}_{-0.05}$ & $-0.049^{+0.035}_{-0.034}$ \\ \\
\hline \\
\label{tab:nsnrun}
\end{tabular}
\end{table}

In the previous section we saw that the inclusion of the VSA data to that
of WMAP shifts the derived limits on the spectral index. Standard,
slow-roll models of inflation predict that the spectral index will be
a function of scale, albeit at a very low level, and it seems a
sensible parameter to allow as the first beyond the standard
model. The analysis of Spergel et al. (2003) and Peiris et al. (2003)
provided evidence for a non-zero value of $n_{\rm
run}(=-0.031^{+0.016}_{-0.017})$ when using CMB data from WMAP, ACBAR
and CBI, along with large-scale structure data from the 2dF galaxy
redshift survey and the Lyman-$\alpha$ forest. This result was
discussed independently by Bridle et al. (2003), Barger et al. (2003),
Leach \& Liddle (2003), Kinney et al. (2003), where it is was shown that
it was highly dependent on the inclusion of the data from the
Lyman-$\alpha$ forest, the veracity of which has been questioned
(Seljak et al. 2003).

We will start our discussion by considering the same model as in the
previous section with no external priors, but with $n_{\rm run}$
allowed to vary. The marginalized distributions and derived limits on $n_{\rm
S}$ and $n_{\rm run}$ are presented in the top row of
Fig.~\ref{fig:nsnrun} and the first three rows of Table~\ref{tab:nsnrun}
for COBE+VSA, WMAP and WMAP+VSA. The derived limits on $\omega_{\rm
b}$, $\omega_{\rm dm}$ and $h$ are not changed appreciably and the other
parameters, $A_{\rm S}$ and $\tau$ (or $z_{\rm re}$) are strongly degenerate and $z_{\rm re}$ will feature in our discussion below.

The values of $n_{\rm S}$ and $n_{\rm run}$ are not particularly well
constrained by COBE+VSA, but it is worth noting that even in this case
there is a definite preference for a value of $n_{\rm run}<0$. The
results have been included for completeness and provide a useful
cross-check.  The results for WMAP are somewhat different to those
presented in Spergel et al. (2003), something to which we will return in
the subsequent discussion. In particular we find that $n_{\rm
run}=-0.060^{+0.037}_{-0.036}$, a $1.6\sigma$ preference for $n_{\rm
run}<0$, as opposed to $n_{\rm run}=-0.047\pm 0.04$ from Spergel et
al. (2003). The significance of this result is improved to $2.2\sigma$
by the inclusion of the high resolution data from the VSA. These
quantitative results are borne out on examination of the likelihood
curves. It is worth emphasizing that this result comes from CMB data
alone.

We have tested the sensitivity of this apparently strong result to the
inclusion of external priors from the HST and 2dF galaxy redshift
survey, and the results are also presented in Fig.~\ref{fig:nsnrun} and
Table~\ref{tab:nsnrun}. We see that the effect of the HST prior is
to relax marginally the constraint on $n_{\rm run}$, although there is a
significant change in the derived limit on $n_{\rm S}$. We note that the
results for WMAP alone are very similar with and without the HST prior.

The inclusion of 2dF does significantly affect our results. Using just
WMAP we find that there is only a marginal preference for $n_{\rm
run}<0$ and the inclusion of VSA only yields a $1.4\sigma$ result. We
note that this is a shift in the derived value and the error bars do
not change significantly; it is worth discussing the reason for this
shift since it is due to the breaking of a degeneracy by the addition
2dF data. The main parameter combination which is constrained by the 
galaxy power
spectrum is the shape parameter $\Gamma=\Omega_{\rm m}h$ which arises
from the size of the horizon at matter-radiation equality measured in
redshift space. Hence, once combined with the CMB data 
the derived parameters $\Omega_{\rm m}$ and
$\Omega_{\Lambda}$ are constrained individually  (Efstathiou et
al. 2002). Fig.~\ref{fig:omolam} presents the marginalized distribution
for these parameters for the three cases: no external prior, HST prior
and 2dF prior. We see that for the first two cases, in which there is
significant evidence for $n_{\rm run}<0$, the preferred values of
$\Omega_{\rm m}$ are much lower (extremely low in the no prior case)
with the corresponding mean values of the distributions giving
$\Omega_{\rm m}h\approx 0.17-0.18$, whereas in the latter case
$\Omega_{\rm m}\approx 0.3$, $h\approx 0.68$ and $\Omega_{\rm
m}h\approx 0.21$, closer to the value suggested by Percival et al. (2001)
from the their analysis of the 2dF alone.

We have also considered the effects of including other CMB information
from the two other high resolution experiments ACBAR and CBI. We find
that the inclusion of their results does not appear to be as
significant as the VSA in preferring a value of $n_{\rm run}<0$ and
that the result of considering WMAP+ACBAR+CBI+VSA is very similar to
just WMAP+VSA. We note that the ACBAR and CBI experiments quote large
global calibration uncertainties ($20\%$ and $10\%$ in power), which
we believe is at least as responsible for this result as their errors
on the individual power spectrum band powers. We note that the calibration
uncertainty for the CBI is likely to improve for future data in a
similar way as for the VSA.  A future re-calibration of CBI and more
data are likely to address this issue (C. Contaldi, priv. comm.).

\subsubsection{Broken power law models}

\begin{figure}
\includegraphics[width=8cm,height=8cm,angle=0]{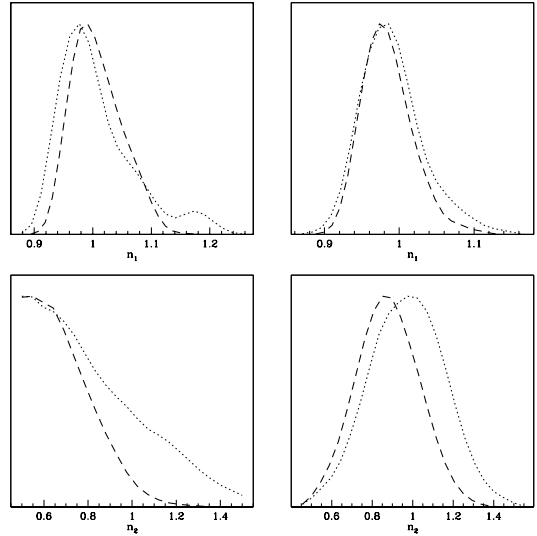}
\caption{Marginalized distributions for $n_1$ and $n_2$ in the
flat $\Lambda$CDM model with a broken power-law index. The line styles
are as in Fig.~\ref{fig:lcdm}. The left-hand column assumes the HST
prior and the right-hand column assumed the 2dF prior.}
\label{fig:n1n2}
\end{figure}

In the previous section we have seen that there is some evidence for
an initial power spectrum of density fluctuations which is not
described by a single power law index. The running spectral index model is
suggested by slow-roll inflation. However, the values which are
preferred by the data, at least with some priors, are too large to come
from standard slow-roll inflation and are incompatible with the idea of the
spectral index being a power series in $\log(k/k_{\rm c})$.  Here, we
consider a model with two spectral indices $n_1$ and $n_2$, with the
cross-over point being $k_{\rm c}=0.05\,{\rm Mpc}^{-1}$.

The results obtained from this model are presented in
Fig.~\ref{fig:n1n2} for the HST and 2dF external priors respectively,
using CMB data from WMAP and WMAP+VSA. We see that, in both cases and
with both datasets, one obtains $n_1\approx 1$. 
The situation for $n_2$ is more
complicated. For the HST case (left column) 
we see that the best fitting value is very
low. In fact, it is  lower than the lower limit we have included as a top hat
prior. For WMAP values as large as $n_2=1.4$ are not
excluded, whereas the inclusion of the VSA has the effect of excluding
models with $n_2>1$. The inclusion of the 2dF prior (right column) 
has a strong effect, moving the distribution of $n_2$ to larger values, but still preferring $n_2<1$. 
In this case for  WMAP we find that
$n_1=0.99\pm 0.04$ and $n_2=0.97\pm 0.18$ which suggests that
something close to scale-invariant $n_1=n_2=1$ is preferred, whereas
for WMAP+VSA $n_1=0.99\pm 0.03$ and $n_2=0.88\pm 0.15$. While this is
clearly compatible with scale invariant even at the $1\sigma$ level,
there is undoubtedly a preference for a broken spectral index
when the VSA is included.

It is clear that our results are compatible with those of the previous
section on running spectral index models. Models with $n_{\rm run}<0$
have a lower value of the spectral index for $k>k_{\rm c}$ than for
$k<k_{\rm c}$ and this is exactly what we find in this alternative
parameterization. We should note that large variations in $n_2$
only lead to much smaller changes in the actual power spectrum than
one might expect naively from, for example, plotting the power
spectrum for different values of $n_{\rm S}$ within the standard
$\Lambda$CDM model.

\subsubsection{Neutrino fraction}

As a final extension to our flat $\Lambda$CDM model, it is of
interest to include the fraction $f_\nu$ of dark matter in the form of
neutrinos. Evidence for a neutrino oscillation, and hence for the
existence of massive neutrinos, has been found by solar neutrino and
atmospheric neutrino experiments (Fukuda et al. 1998, 2002; Allison et
al. 1999; Ambrosio et al. 2000; Ahmad et al. 2002). Further evidence
for a non-zero value of the neutrino mass has recently been claimed
from cosmological data (Allen et al. 2003b).

In addition to obtaining constraints on $f_\nu$, the inclusion of this
parameter will inevitably lead to some broadening of the marginalized
distributions for the other parameters. Of particular interest is
whether the constraints on the running spectral index derived above
are robust to the inclusion of $f_\nu$. We therefore include $f_\nu$,
with the top-hat prior given in Table~\ref{tab:pri}, into the running
spectral index model. In the analysis of this model, we include the
2dF external prior, since current CMB alone provide only a weak 
constraint on $f_\nu$.

\begin{figure*}
\includegraphics[width=0.3\linewidth,angle=0]{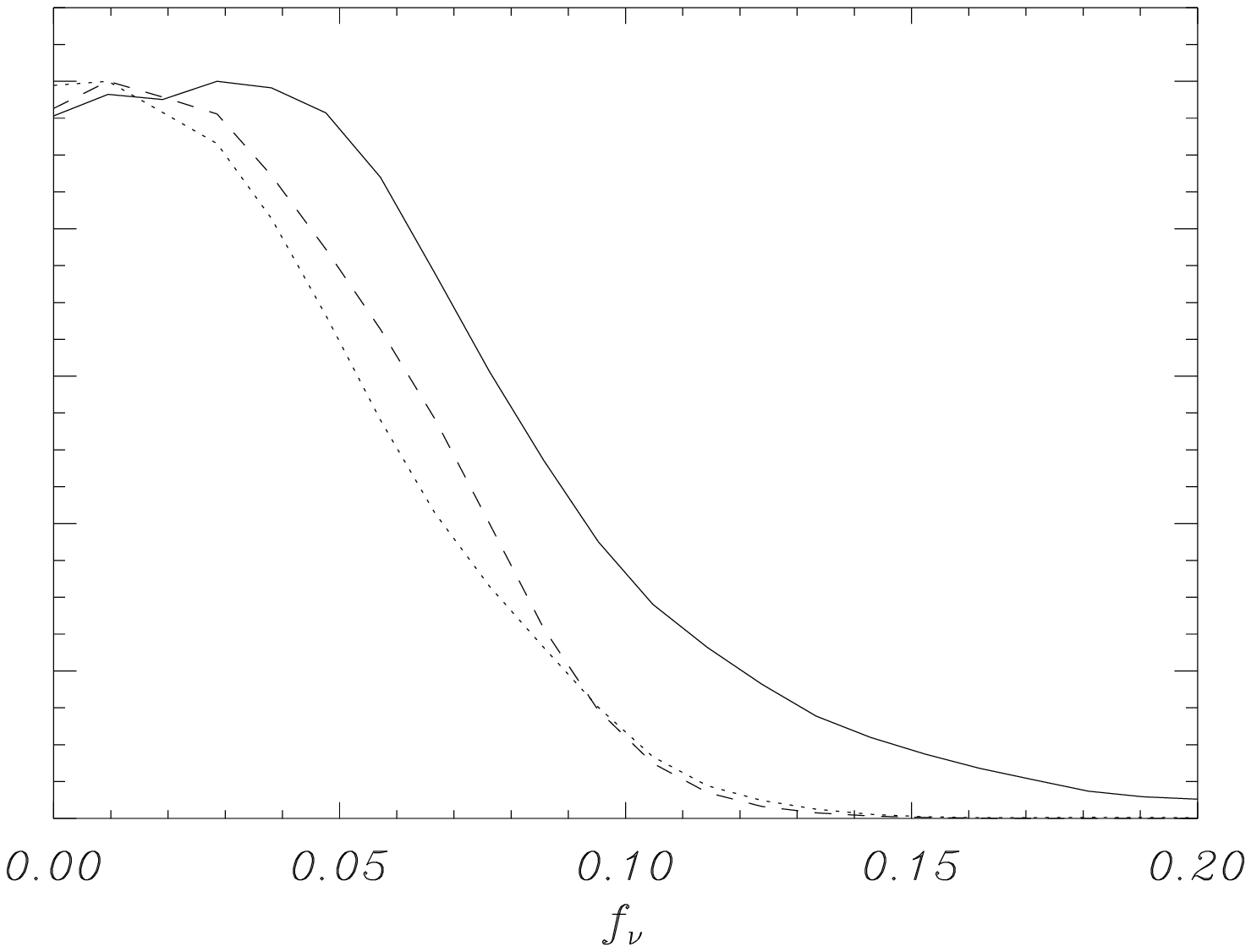}
\includegraphics[width=0.3\linewidth,angle=0]{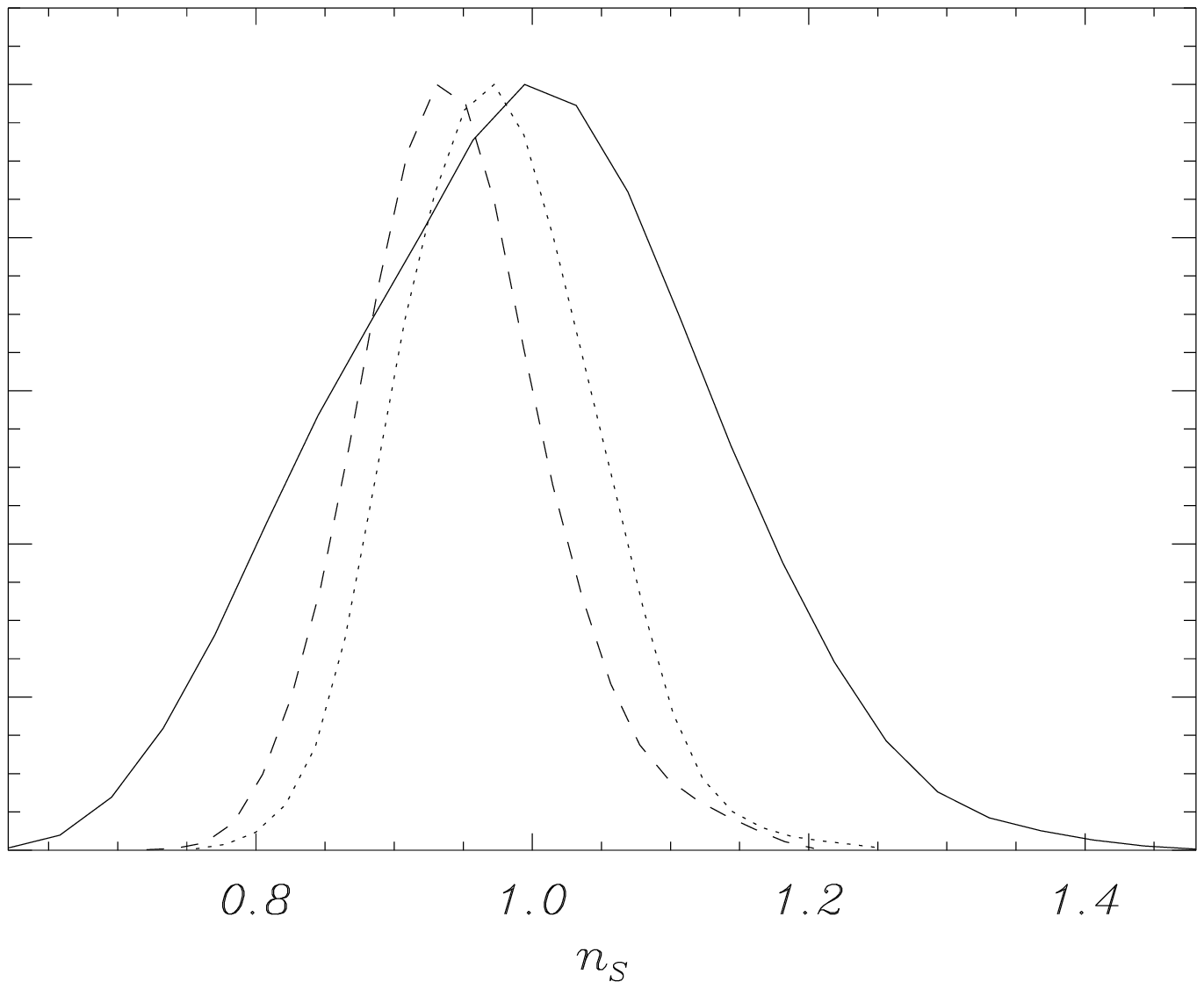}
\includegraphics[width=0.3\linewidth,angle=0]{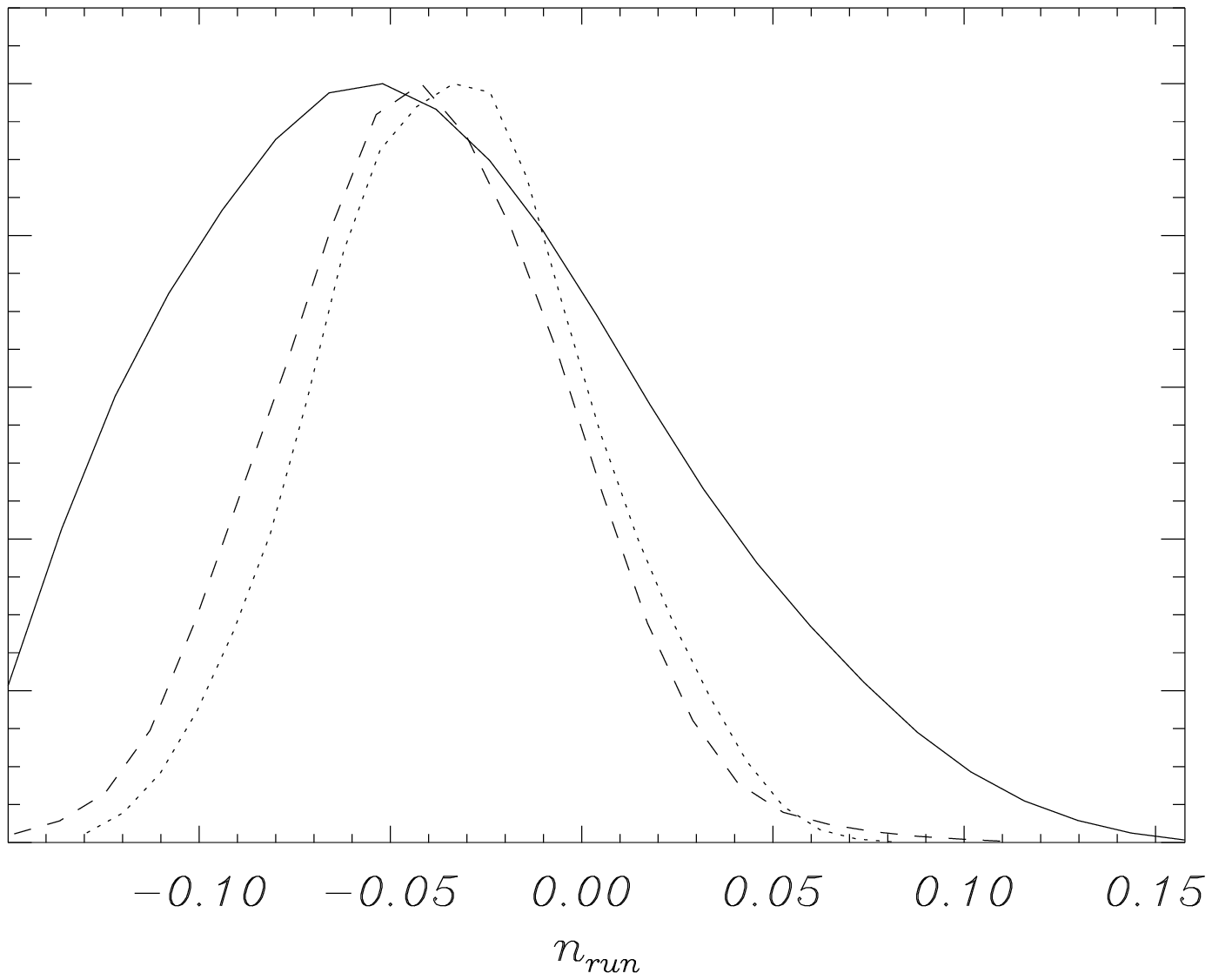}
\caption{Marginalized distributions for 
$f_\nu$, $n_s$ and $n_{\rm run}$ in 
the extended flat $\Lambda$CDM model, using the 2dF external prior and 
COBE+VSA (solid-line), WMAP alone (dotted line) and WMAP+VSA (dashed
line).}
\label{fig:fnu_lcdm}
\end{figure*}
 
Fig.~\ref{fig:fnu_lcdm} shows the marginalized distributions
obtained for $f_\nu$, $n_{\rm S}$ and $n_{\rm run}$ for the three CMB data
sets COBE+VSA, WMAP and WMAP+VSA.  We find that the 95\% upper limit
provided by the COBE+VSA data set, $f_\nu < 0.132$, is only marginally
larger than that obtained using WMAP data, $f_\nu < 0.090$.  The
combination WMAP+VSA gives similar limits to WMAP, namely $f_\nu <
0.087$, which corresponds to neutrino mass of
$m_\nu<0.32{\rm eV}$ when the neutrino masses are degenerate.

For the parameters $n_{\rm S}$ and $n_{\rm run}$, we see that, as compared
with those plotted in Fig.~\ref{fig:nsnrun} (middle row), the
marginalized distributions have indeed been shifted and 
broadened by the inclusion
of $f_\nu$ although the effects are not very strong.
In particular, we note that our earlier finding of a
preference for a non-zero value of $n_{\rm run}$ has been
weakened somewhat. A non-zero $n_{\rm
run}$ is still preferred, but at reduced significance.
For the WMAP+VSA data set, we obtain $n_{\rm S} =
0.94_{-0.06}^{+0.06}$ and $n_{\rm run} = -0.041_{-0.036}^{+0.037}$
with 68\% confidence limits.

\begin{figure*}
\includegraphics[angle=-90,width=15cm]{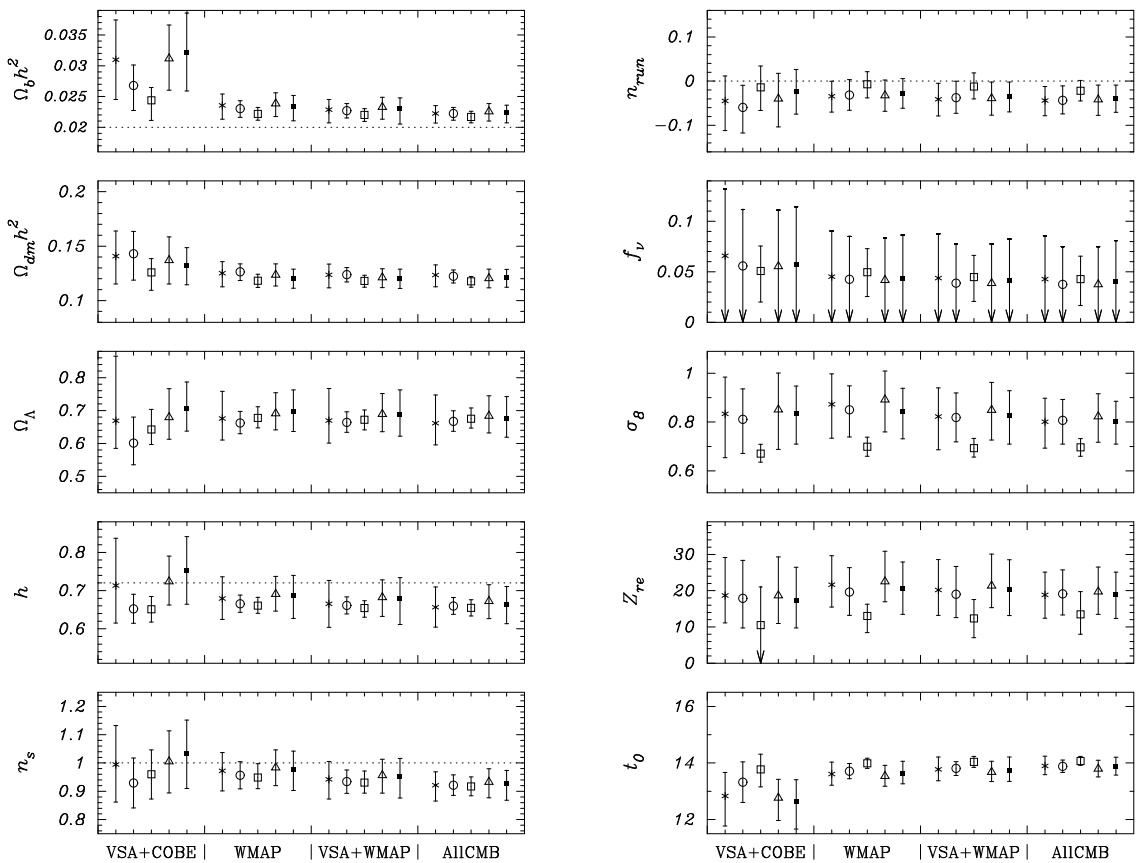}
\medskip
\caption{Estimates for cosmological parameters in the flat
$\Lambda$CDM running spectral index model, extended to include
$f_\nu$. Four CMB data sets are considered and, for each data set,
four determinations are plotted, corresponding to different
combinations of external priors. From left to right the external
priors are: 2dF; 2dF+$f_{\rm gas}$; 2dF+$f_{\rm gas}$+XLF;
2dF+HST and 2dF+CS.
The points indicate the median of the corresponding marginal
distributions. The error bars denote 68\% confidence limits.  If a
distribution peaks at zero then the
95\% upper limit is shown. The horizontal dashed lines plotted in some
of the panels indicate BBN values for $\Omega_{\rm b} h^2$, the value
of $h$ given by the HST key project, the Harrison-Zel'dovich value
of the spectral index of fluctuations and a zero value for the running
index.}
\label{fig:flatis}
\end{figure*}

In the above analysis we used only 2dF as an external prior. It is of
interest to investigate the effect of including different combinations
of the additional external priors listed in
Table~\ref{tab:pri}. The effect of these additional priors has
been calculated by importance sampling our previous results. We also
investigate the effect of including all recent CMB data into our
analysis. In Fig.~\ref{fig:flatis}, we plot confidence limits on all
the model parameters for each of our four CMB data sets, each of
which, in turn, includes four different combinations of external
priors: 2dF, 2dF+$f_{\rm gas}$, 2dF+$f_{\rm gas}$+XLF, 2df+HST and 2dF+CS. The
points indicate the median of the corresponding marginalized
distribution, and the error bars show the 68\% central confidence
limit. If the distribution peaks at zero, the point is placed on the
axis and the 95\% upper limit is shown.
\begin{table}
\caption{Parameter estimates and 68\% confidence intervals for various
cosmological parameters as derived from Fig.~\ref{fig:general}. For
$f_\nu$ and $R$, the 95\% upper limits are quoted.}
\begin{tabular}{lrrr}
\hline

 &WMAP & WMAP+VSA & {\sc AllCMB} \\
\hline
\\
$\Omega_{\rm b} h^2$
 &  $	 0.025  ^{+0.003}_{-0.003} $
 &  $	 0.024  ^{+0.003}_{-0.002} $
 &  $	 0.023  ^{+0.002}_{-0.002} $  \\ \\
$\Omega_{\rm dm} h^2$
 &  $	 0.108  ^{+0.022}_{-0.021} $
 &  $	 0.111  ^{+0.021}_{-0.019} $
 &  $	 0.113  ^{+0.017}_{-0.017} $  \\ \\
$h$
 &  $	0.66  ^{+0.07}_{-0.06} $
 &  $	0.66  ^{+0.06}_{-0.06} $
 &  $	0.65  ^{+0.07}_{-0.07} $  \\ \\
$z_{\rm re}$
 &  $	18  ^{+7}_{-7} $
 &  $	19  ^{+7}_{-7} $
 &  $	17  ^{+7}_{-8} $  \\ \\
$\Omega_{\rm k}$
 &  $	 -0.02  ^{+0.03}_{-0.03} $
 &  $	 -0.01  ^{+0.03}_{-0.03} $
 &  $	 -0.02  ^{+0.03}_{-0.03} $  \\ \\
$f_\nu$
 &  $	 < 0.093 $
 &  $	 < 0.083 $
 &  $	 < 0.083 $   \\ \\
$w$
 &  $	 -1.00  ^{+0.24}_{-0.27	   } $
 &  $	 -0.99  ^{+0.24}_{-0.27	   } $
 &  $	 -1.06  ^{+0.24}_{-0.25	   } $  \\ \\
$n_{\rm S}$
 &  $	 1.04  ^{+0.12}_{-0.11} $
 &  $	 0.99  ^{+0.09}_{-0.09} $
 &  $	 0.96  ^{+0.07}_{-0.07} $  \\ \\
$n_{\rm T}$
 &  $	 0.26  ^{+0.53}_{-0.60} $
 &  $	 0.13  ^{+0.49}_{-0.51} $
 &  $	 0.12  ^{+0.48}_{-0.51} $  \\ \\
$n_{\rm run}$
 &  $	 -0.02  ^{+0.07}_{-0.05} $
 &  $	 -0.04  ^{+0.05}_{-0.04} $
 &  $	 -0.04  ^{+0.04}_{-0.05} $  \\ \\
$10^{10} A_{\rm S}$
 &  $	27  ^{+8}_{-5} $
 &  $	26  ^{+9}_{-5} $
 &  $	25  ^{+6}_{-5} $  \\ \\
$R$
 &  $	 < 0.78 $
 &  $	 < 0.77 $
 &  $	 < 0.68 $  \\ \\
$\Omega_\Lambda$
 &  $	 0.71  ^{+0.07}_{-0.09	   } $
 &  $	 0.70  ^{+0.06}_{-0.08	   } $
 &  $	 0.69  ^{+0.07}_{-0.09	   } $  \\ \\
$t_0$
 &  $	14.1  ^{+1.4}_{-1.1} $
 &  $	14.1  ^{+1.3}_{-1.2} $
 &  $	14.4  ^{+1.4}_{-1.3} $  \\ \\
$\Omega_{\rm m}$
 &  $	 0.31  ^{+0.09}_{-0.07} $
 &  $	 0.31  ^{+0.08}_{-0.06} $
 &  $	 0.33  ^{+0.10}_{-0.07} $  \\ \\
$\sigma_8$
 &  $	 0.76  ^{+0.14}_{-0.14} $
 &  $	 0.77  ^{+0.13}_{-0.13} $
 &  $	 0.76  ^{+0.11}_{-0.12} $  \\ \\
$\tau$
 &  $	 0.20  ^{+0.13}_{-0.11} $
 &  $	 0.20  ^{+0.15}_{-0.10} $
 &  $	 0.17  ^{+0.12}_{-0.10} $  \\ \\
\hline
\label{tab:general}
\end{tabular}
\end{table}

We see that the inclusion of the $f_{\rm gas}$ and XLF external priors
significantly reduces the error bars on all parameters. The most
profound effect is obtained from the XLF prior for the parameters
$f_\nu$, $\sigma_8$ and $z_{\rm re}$, as might be expected from Allen
et al. (2003b). Indeed, it is only with the inclusion of the XLF prior
that a non-zero value of $f_\nu$ is preferred and only then at limited significance. For each of the CMB 
data set combinations, the best-fitting value in this case is
$f_\nu\approx 0.05$, which corresponds to neutrino mass of $m_\nu
\approx 0.18{\rm eV}$ when the neutrino masses are degenerate, with a
zero value excluded at around 96\% confidence.  For $\sigma_8$ the
inclusion of the XLF prior significantly reduces the best-fit value
and the error bars for all CMB data set combinations. A similar, but
less pronounced, effect is seen for $z_{\rm re}$.

\subsection{General $\Lambda$CDM model}

Thus far we have considered only a limited range of flat $\Lambda$CDM models.
In principle, one should properly include all the relevant unknowns
into the analysis in order to obtain conservative confidence
limits. In this section, we consider a more general
$\Lambda$CDM model. In addition to including $f_\nu$ and $n_{\rm run}$, the
standard six-parameter flat $\Lambda$CDM model is further extended by
including $\Omega_k$, $w$, $R=A_{\rm T}/A_{\rm S}$ and $n_{\rm T}$. This
gives 12 variable parameters in total, for which we adopt the top-hat priors
listed in Table~\ref{tab:pri}.

\begin{figure*}
\includegraphics[width=17cm,angle=0]{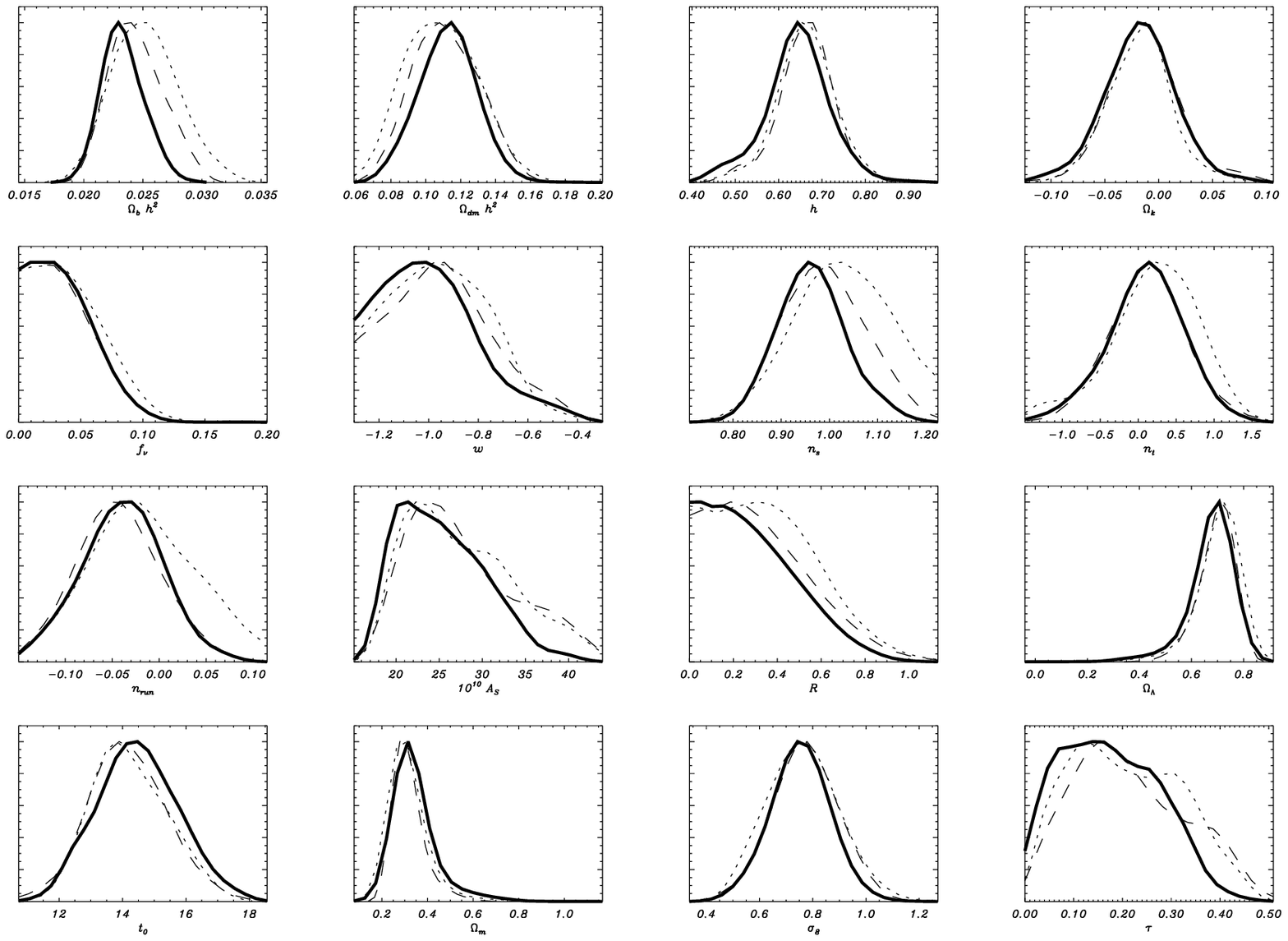}
\caption{Marginalized distributions for various cosmological
parameters in the 12-parameter general non-flat $\Lambda$CDM model from 
WMAP (dotted line), WMAP+VSA (dashed line) and {\sc AllCMB} (thick solid
line) in combination with external priors from both 2dF and SNeIa.}
\label{fig:general}
\end{figure*}

For this model, we consider the three CMB data sets WMAP, WMAP+VSA and
{\sc AllCMB}. In addition, we now use both 2dF and SNeIa as our basic
 external priors, which are required in order to set constraints
on our 12-dimensional cosmological parameter space. For each CMB data
set, the marginalized distributions for each parameter are shown in
Fig.~\ref{fig:general}.  In addition, marginalized distributions are
plotted for the derived parameters $\Omega_\Lambda$, $\Omega_{\rm m}$,
$t_0$ and $\sigma_8$. The corresponding confidence limits on the
parameter values are given in Table~\ref{tab:general}.

In this more general model we see that the marginalized distributions of
the parameters in our simpler models have broadened somewhat, but are
still consistent with our earlier findings. Perhaps most interesting
is the fact that some of the marginalized distributions change
considerably as more CMB data are used beyond WMAP.  For $\Omega_{\rm
b}h^2$ we see a clear trend towards a lower preferred value (closer to
the BBN estimate) as one adds first VSA data and then all remaining
CMB data sets. This effect is accompanied by a gradual upwards trend in
the preferred $\Omega_{\rm dm}h^2$ value. The other parameters
exhibiting such trends are $n_{\rm S}$ and $n_{\rm run}$. As more CMB data
are included, the preferred value of $n_{\rm S}$ moves slightly below unity,
although this value is by no means excluded. Perhaps more importantly, the
upper limit on $n_{\rm S}$ is significantly reduced as more CMB data are
added. An analogous effect is observed for $n_{\rm run}$, for which
the addition of VSA data significantly reduces the tail of the
distribution for positive values of $n_{\rm run}$.

The remaining marginalized distributions have very similar forms for each
of the three CMB data sets, indicating that, for these parameters, the
addition of further CMB beyond WMAP does not have a significant effect
on their derived values and confidence limits. It is worth noting in
passing, however, that all CMB data sets are fully consistent with a
zero curvature model. Moreover, we find $w= -1$ with an uncertainty of
$\pm$ 24\%, which is consistent with dark energy in the form of a
cosmological constant.  As regards inflation models, we find that the
inclusion of VSA data makes a modest reduction in the upper limit on
the tensor-to-scalar ratio, which is reduced still further (albeit
marginally) by the inclusion of all CMB data; in this last case we
obtain $R < 0.68$ at 95\% confidence. The power-law index of tensor
modes $n_{\rm T}$ is fully consistent with zero.

\begin{figure*}
\includegraphics[angle=-90,width=15cm]{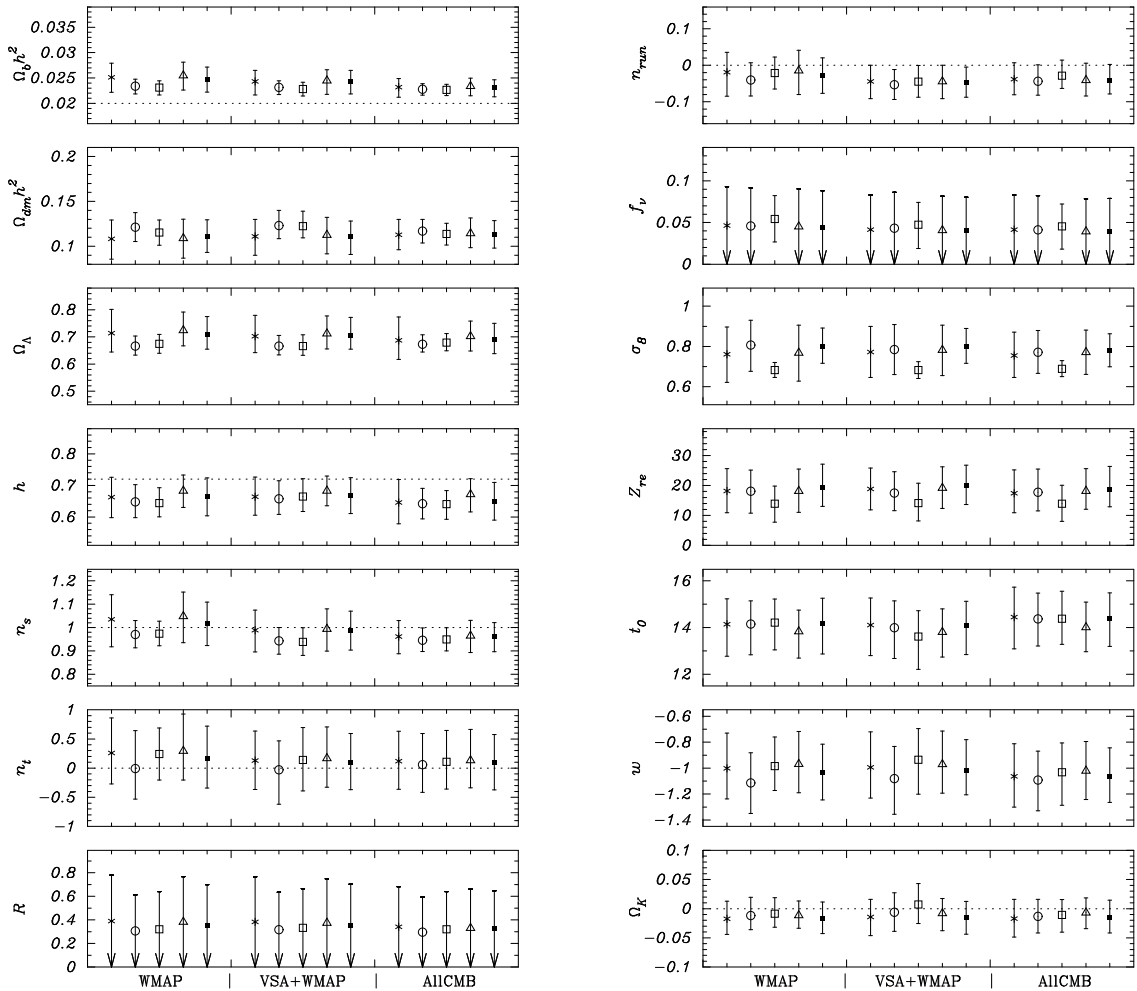}
\medskip
\caption{As for Fig.~\ref{fig:flatis}, but for the general
12-parameter non-flat $\Lambda$CDM model.}
\label{fig:generalis}
\end{figure*}

As we did for the flat $\Lambda$CDM model, we may investigate the
effect of including additional external priors in our analysis of the
general model. In Fig.~\ref{fig:generalis} we plot the 
confidence intervals on all the model parameters for each of our four
CMB data sets, each of which, in turn, includes four different
combinations of external priors: 2dF, 2dF+$f_{\rm gas}$, 2dF+$f_{\rm
gas}$+XLF, 2dF+HST and 2dF+CS. Once again, we see that the inclusion of
the $f_{\rm gas}$ and XLF external priors has the greatest effect on
the confidence limits, and that this is most pronounced for the XLF
prior and the parameters $f_\nu$, $\sigma_8$ and $z_{\rm re}$. It is
reassuring, however, that the derived limits on $f_\nu$ for the
general model are very similar to those obtained assuming the simpler
flat model. We again find $f_\nu \approx 0.05$, with a zero-value
excluded at about 92\% confidence which is slightly lower than for the flat case.  The effect of the XLF prior on
$\sigma_8$ and $z_{\rm re}$ in the general model is also similar to
that observed in the simpler flat case.

\section{Discussion and conclusions}
\label{sec:disc}

We have used recent data from the Very Small Array, together with
other CMB datasets and external priors, to set constraints on
cosmological parameters. We have considered both flat and non-flat
$\Lambda$CDM models and the results are consistent.

Within the flat $\Lambda$CDM model, we find that the inclusion of VSA
data suggests that the initial fluctuation spectrum that is not
described by a single power-law. As we have pointed out already, the
value of $n_{\rm run}$ preferred by the data is incompatible with the
basic premises of slow-roll inflation. Moreover, the negative running, which reduces the amount of power on small scales and hence the amount of structure at early times,
leads to predictions for the epoch of reionization at odds with the best fit to the CMB data. This comes almost
directly from the temperature-polarization 
cross-correlation power spectrum observed by WMAP
(Kogut et al. 2003). Given the implications of this result it is
important to consider the possible systematic effects that might
weaken it.

The absolute calibration uncertainty of the VSA power spectrum is an important
contributory factor to this result. The 3\% uncertainty quoted in
Dickinson et al. (2004)
relies heavily on the the measurement of the temperature of
Jupiter, $T_{\rm jup}$, by WMAP and this requires an overall factor of
0.92 modification in the power spectrum
estimates from the previous VSA results (Grainge et al 2003; Scott et
al 2002) which were reliant on on earlier measurements of $T_{\rm
jup}$ given 
by Mason et al. (1999). It was pointed out in Dickinson et al. that, in
fact, using an absolute calibration based on this measurement of
$T_{\rm jup}$ gives the most consistent normalization of the power
spectrum when compared to that of WMAP.

We have investigated the effects of possible uncertainties in the
calibration in two ways. First, we consider the possibility of using
the Mason et al (1999) central value for
$T_{\rm jup}$ while maintaining an overall
uncertainty of $3\%$. As an alternative we just increase the overall
uncertainty in the calibration to $10\%$ while keeping the central value for $T_{\rm jup}$ from WMAP. The derived limits on
$n_{\rm S}$ and $n_{\rm run}$ are presented in Table~\ref{tab:calib}
for these two possibilities using the HST and 2dF priors. We see in
each case that the preference for $n_{\rm run}<0$ is weakened to below
$2\sigma$ compared to the calibration based on WMAP's measurement of
$T_{\rm jup}$.  It is clear that refinement of the absolute
calibration of the VSA in the light of the WMAP measurements is
something requiring further attention.

\begin{table*}
\caption{Limits on $n_{\rm S}$ and $n_{\rm run}$ in the flat
$\Lambda$CDM model with a running spectral index for different
absolute calibration schemes. The uncertainty refers to that in the power. See text for discussion.}
\begin{tabular}{lcccc}
\hline
$T_{\rm jup}$ & Uncertainty  & External & $n_{\rm S}$ & $n_{\rm run}$ \\
\hline \\
Mason et al. & 3\% & HST & $0.93^{+0.05}_{-0.05}$ & $-0.058^{+0.038}_{-0.038}$ \\ \\
Mason et al. & 3\% & 2dF & $0.93^{+0.05}_{-0.05}$ & $-0.028^{+0.047}_{-0.047}$ \\ \\ \\
WMAP & 10\% & HST & $0.93^{+0.05}_{-0.05}$ & $-0.055^{+0.035}_{-0.035}$ \\ \\
WMAP & 10\% & 2dF & $0.95^{+0.06}_{-0.06}$ & $-0.040^{+0.033}_{-0.033}$ \\ \\
\hline\\
\label{tab:calib}
\end{tabular}
\end{table*}

Another possible systematic effect is the residual point
source correction due to sources below out subtraction limit of
$20{\rm mJy}$.
This was computed by normalizing
the point source model of Toffolatti et al. (1998) to the observed VSA
source counts which can then be extrapolated to lower flux
densities. There are clearly some uncertainties in this procedure. It
is possible that an imperfect subtraction, either an over-estimate or
an under-estimate, could lead to inaccuracies in the derived limits on
the cosmological parameters, in particular on $n_{\rm S}$ and $n_{\rm
run}$. In order to investigate possible effects of such uncertainties
we have performed our likelihood analysis with the inclusion of the
parameter $A_{\rm X}$ which was discussed in section~\ref{sec:cosmomodel}. We note that it is also possible that for Galactic foregrounds might contribute to this. However, it was shown in Dickinson et al. (2004) that the level of foreground contamination of the VSA fields was negligible.

\begin{figure}
\includegraphics[width=8cm,height=12cm,angle=0]{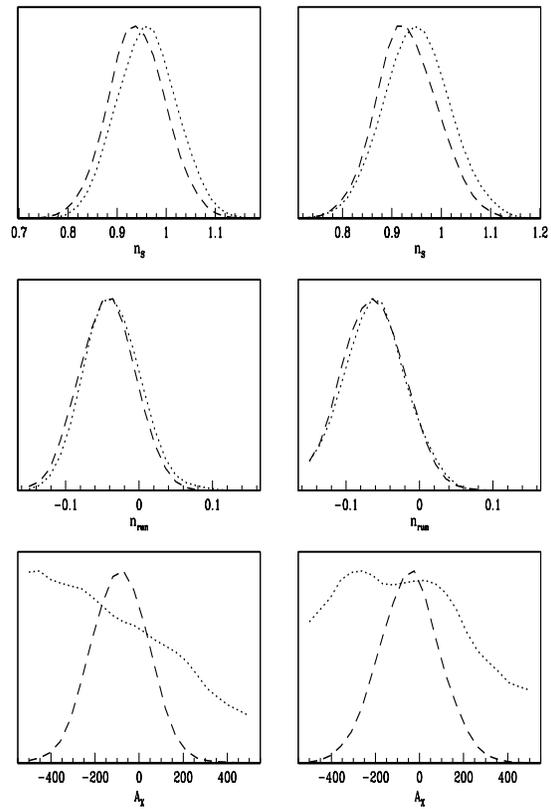}
\caption{Marginalized distributions for $n_{\rm S}$, $n_{\rm run}$ and $A_{\rm X}$ when $A_{\rm X}$ is allowed to vary. The line-styles are as in Fig.~\ref{fig:lcdm}. but COBE+VSA is excluded The left-hand column is for the HST prior and the right-hand for the 2dF prior.}
\label{fig:ax}
\end{figure}

We find that the derived limits on $n_{\rm S}$ and $n_{\rm run}$ in
this case are less stringent than without including $A_{\rm X}$ for WMAP+VSA.
The marginalized distributions for $n_{\rm S}$, $n_{\rm run}$ and $A_{\rm
X}$ are presented using CMB data from WMAP and WMAP+VSA for the
external priors from HST and 2dF in Fig.~\ref{fig:ax} and the derived
limits are presented in Table.~\ref{tab:ax}. In fact the likelihood curves and derived limits for WMAP and WMAP+VSA are almost identical when $A_{\rm X}$ is included in the analysis and the WMAP limits are very similar to when $A_{\rm X}$ is constrained to be  zero (see Table~\ref{tab:nsnrun}). We see that there are essentially no limits on $A_{\rm X}$ when just considering WMAP and that in the case of WMAP+VSA, $A_{\rm X}$ is compatible with zero suggesting that to within at least  $\approx 100(\mu{\rm K})^2$ the source subtraction procedure has been successful. For the HST prior $A_{\rm X}<214\,(\mu{\rm K})^2$ at 95\% confidence and for the 2dF prior $A_{\rm X}<155\,(\mu{\rm K})^2$.

\begin{table*}
\caption{Limits on $n_{\rm S}$, $n_{\rm run}$ and $A_{\rm X}$ in the flat
$\Lambda$CDM model with a running spectral index when the parameter $A_{\rm X}$ is included. The final column is the 95\% confidence upper limit on $A_{\rm X}$. The units of $A_{\rm X}$ are in $(\mu K)^2$.}
\begin{tabular}{lccccc}
\hline
CMB & External & $n_{\rm S}$ & $n_{\rm run}$ & $A_{\rm X}$ & $A_{\rm X}\,(2\sigma)$ \\
\hline \\
WMAP &  HST & $0.95^{+0.06}_{-0.06}$ & $-0.059^{+0.039}_{-0.039}$ & Unconstrained & \\ \\
WMAP+VSA &HST & $0.93^{+0.06}_{-0.06}$ & $-0.061^{+0.038}_{-0.038}$ & $-46\pm 132$ & $<214$  \\ \\ \\
WMAP &  2dF & $0.96^{+0.06}_{-0.06}$ & $-0.036^{+0.036}_{-0.036}$ & Unconstrained & \\ \\
WMAP+VSA & 2dF & $0.94^{+0.06}_{-0.06}$ & $-0.043^{+0.035}_{-0.035}$ & $-86\pm 123$ & $<155$  \\ \\
\hline \\
\label{tab:ax}
\end{tabular}
\end{table*}

Assuming that the subtraction is perfect and that the SZ contribution to the power spectrum is $\propto \ell^2$ for $\ell<4000$ then our results on $A_{\rm X}$ can be used to derive a limit on $(\Delta T_\ell)^2$ in a bandpower, $B_{3000}$,  covering $2000\le\ell\le 4000$ as observed by CBI. Under the assumption that $\Delta T_{\ell}\propto \ell^2$, we find that 
\begin{equation}
B_{3000}={28\over 3}A_{\rm X}\,.
\label{b3000}
\end{equation}
This leads to a
limit of $B_{3000}< 1997\,(\mu{\rm K})^2$ using the HST prior and $B_{3000}< 1446\,(\mu{\rm K})^2$ for the 2dF prior. The value quoted by Mason et al. (2003) is $B_{3000}=(508\pm 168)\,(\mu{\rm K})^2$ which is more stringent than our limit. However, this value and our limit are derived in very different ways. The measurement of Mason et al. (2003) is a direct limit from high resolution imaging 
of 3 deep fields. While it is direct, the measurement the global power spectrum could be significantly affected by sample variance; one could have observed a field which in there are more clusters than the global average and hence obtain a biased estimation of the global power spectrum.
Our limit is indirect, coming from the power spectrum measured
over 82~${\rm deg}^2$ at lower angular resolution and it requires that the power spectrum be $\propto\ell^2$ as well as the cosmological model to be correct. It is likely that this represents a reliable upper bound since the power spectrum will grow less rapidly than $\ell^2$ for $\ell>1500$ due to the fact that the clusters responsible for the SZ effect are not point sources. Moreover, it is not as sensitive to the Poisson distributed  number of clusters in an individual field. A more realistic modelling of the SZ effect using accurate power spectra could yield a more stringent upper bound. The two methods provide useful complimentary information and it is possible that a much more stringent constraint on $A_{\rm X}$ will be possible when the VSA observes with higher resolution in the near future.

One important feature of the power spectrum observed by WMAP
is the apparent absence of power at very 
low $\ell$. This could be due to some as
yet unknown physics, or it could be a manifestation of the interaction between the subtle systematic effects caused by the
side-lobes, the Galactic cut, and the power spectrum estimation
algorithm used by the WMAP team (Efstathiou 2003). It is worth
assessing to what extent our result is dependent on the measured
anisotropies with $\ell<10$. By excluding the multipoles with $\ell<10$ from our analysis we find that $n_{\rm S}=1.01\pm 0.07$ and $n_{\rm run}=0.007\pm 0.049$, strongly suggestive of an $n_S\equiv 1$, scale invariant initial power spectrum for WMAP with the 2dF prior, whereas $n_{\rm S}=0.97\pm 0.06$ and $n_{\rm run}=-0.015\pm 0.047$ for WMAP+VSA and the same prior. The weakening of the constraint on $n_{\rm run}$ should not be a surprise since excluding multipoles with $\ell<10$ cuts out nearly a whole power of ten in $k$ and $n_{\rm run}$ is the coefficient of a power series in $\log(k/k_{\rm c})$. However, we see that the inclusion of the VSA tends to prefer a spectral index lower than just WMAP. It is clear from this that the reason for the preference for a negative running of the power spectrum when multipoles with $\ell<10$ are included is the tension between the measurements at $\ell<10$ by WMAP and for $\ell>1000$ by the VSA.

We should comment briefly on one aspect of our analysis of the running spectral index models which is not ideal: the preferred values of $z_{\rm re}$ and $\tau$. Most recent analyses of CMB data include an upper bound one of these parameters. In Spergel et al. (2003) a flat prior of $\tau<0.3$ was used when in some cases the data had a preference for a high value of $\tau$ by virtue of the low-$\ell$ TE correlation power spectrum; our analysis is no different and we believe that this is responsible for the differences between our analysis and that of Spergel et al (2003). All the likelihood curves and derived limits have made the not unreasonable assumption that $z_{\rm re}<30$. However, in some cases, particularly those for which we have included no prior from 2dF, the preferred values of $z_{\rm re}$ are close to this limit, uncomfortably close in some cases and one might be concerned that our results are sensitive to this. We find that the models with $n_{\rm run}$ significantly less than zero tend to have larger values of $z_{\rm re}$ which explains why our derived limit of $n_{\rm run}=-0.060\pm 0.037$ from WMAP is larger than the one quoted in Spergel et al (2003). It also suggests that, by excluding $z_{\rm re}>30$, we have weakened the constraint on $n_{\rm run}$ rather than artificially modifying the preferred value away from zero. An epoch of reionization with $z_{\rm re}\approx 30$ would seem unlikely in the context of early structures being the source of ionization, but it is clear that the data suggest it.

For the general 12-parameter $\Lambda$CDM model, we find that our
marginalized distributions for $n_{\rm S}$ and $n_{\rm run}$ are
broadened, as one would expect. Nevertheless, even in this case, the
addition of VSA data significantly reduces tails of the distributions
for $n_{\rm S}$ greater than unity and for positive $n_{\rm
run}$, as compared with using WMAP as the only CMB data set.
Indeed, these effects are reinforced by the use of the {\sc AllCMB}
data set. The inclusion of additional CMB data beyond WMAP also leads
to a noticeable reduction in the preferred value of $\omega_{\rm b}$
and a corresponding increase in $\omega_{\rm dm}$.

To summarize, we find that there is evidence for $n_{\rm run}<0$ in a limited class of models, but within the general $\Lambda$CDM model with 12 parameters the evidence is much weaker. Standard models of inflation are generally incompatible with such large negative values of $n_{\rm run}$, but the data appears to point in that direction, although not totally conclusively. The inclusion of an external prior from 2dF appears to weaken the result by fixing $\Omega_{\rm m}\approx 0.3$ in conjunction with the CMB data. The measurement of $\Omega_{\rm m}h$ using the galaxy power spectrum is responsible for this shift. It is an interesting question as to how reliable this measurement is since a slight shift in the results, a preference for $\Omega_{\rm m}h\approx 0.17$ rather than $\Omega_{\rm m}h\approx 0.21$ would bring their preferred value into line with that suggested by the CMB alone and would uphold the possibility of $n_{\rm run}<0$. Since none of the galaxy redshift surveys have conclusively observed the turnover in the power spectrum on which this determination of $\Omega_{\rm m}h$ is based we assert that there is still room for some doubt. We have also investigated the possible systematic effects which could weaken our result. We believe that the assumptions behind the power spectrum measurements presented in Dickinson et al. (2004) are the best available using the observations which we have made and the other information from the literature we have used. For sure, we have shown that measurements of the CMB power spectrum beyond $\ell=1000$ can have an impact on the estimation of cosmological parameters and that future measurements in this region by the VSA, the PLANCK satellite and others will enable us in the future to make more definitive statements.

\section*{ACKNOWLEDGEMENTS} 

We acknowledge S. Allen for permission to use his XLF probability
curves, and H. Hoekstra for providing probability curves for cosmic
shear. JAR-M acknowledges the hospitality of the IAC during several
visits, and the financial support provided through the European
Community's Human Potential Programme under contract
HPRN-CT-2002-00124, CMBNET. KL, RS, CD acknowledge support by PPARC studentships. YAH is supported by the Space Research Institute of KACST. AS acknowledges the support of St. Johns College Cambridge. RAB thanks C. Contaldi and J. Weller for helpful comments and suggestions, and acknowledges the use of the COSMOS supercomputer based at DAMTP, Cambridge.

\label{lastpage}
%\bibliography{../cmb_refs}
\bibliographystyle{mn2e}

\bsp
\end{document}